# Stain Normalized Breast Histopathology Image Recognition using Convolutional Neural Networks for Cancer Detection

Sruthi Krishna[1], Suganthi S.S[2], Shivsubramani Krishnamoorthy[3], Arnav Bhavsar[4]

Center for Wireless Networks & Applications (WNA), Amrita Vishwa Vidyapeetham, Amritapuri, India[1]
Valeo, Navalur, India[2]
Department of Computer Science and Engineering, Amrita Vishwa Vidyapeetham, Amritapuri, India[3]
School of Computing and Electrical Engineering, IIT Mandi, India[4]
*sruthik@am.amrita.edu

**Abstract**

**Background and objectives:** Computer assisted diagnosis in digital pathology is becoming ubiquitous as it can provide more efficient and objective healthcare diagnostics. Recent advances have shown that the convolutional Neural Network (CNN) architectures, a well-established deep learning paradigm, can be used to design a Computer Aided Diagnostic (CAD) System for breast cancer detection. However, the challenges due to stain variability and the effect of stain normalization with such deep learning frameworks are yet to be well explored. Moreover, performance analysis with arguably more efficient network models, which may be important for high throughput screening, is also not well explored.

**Methods:** To address this challenge, we consider some contemporary CNN models for binary classification of breast histopathology images that involves (1) the data preprocessing with stain normalized images using an adaptive colour deconvolution (ACD) based color normalization algorithm to handle the stain variabilities; and (2) applying transfer learning based training of some arguably more efficient CNN models, namely Visual Geometry Group Network (VGG16), MobileNet and EfficientNet. We have validated the trained CNN networks on a publicly available BreaKHis dataset, for 200x and 400x magnified histopathology images.

**Results:** The experimental analysis shows that pretrained networks in most cases yield better quality results on data augmented breast histopathology images with stain normalization, than the case without stain normalization. Further, we evaluated the performance and efficiency of popular lightweight networks using stain normalized images and found that EfficientNet outperforms VGG16 and MobileNet in terms of test accuracy and F1 Score. We observed that efficiency in terms of test time is better in EfficientNet than other networks; VGG Net, MobileNet, without much drop in the classification accuracy.

**Conclusion:** We believe that the study of classification models in this work for breast cancer diagnosis will aid in providing further directions for developing a reliable and efficient system to assist pathologists.

1. **Introduction**

Breast Cancer is one of the most common cancer affecting women. It is estimated that in 2018 alone more than 1,00,000 new cases of breast cancer were diagnosed in India [1]. The consequent mortality rates are notably high in developing countries, supervened by inadequate outfitted facilities for timely diagnosis and treatment. Although numerous imaging modalities aiding the diagnosis of breast cancer, the gold standard validation of cancer diagnosis involves histopathology observation to breast biopsies. Histopathology refers to microscopic examination of tissue mounted slides to analyse the structure of the nuclei, cytoplasm, and stroma in the body tissues. The analysis of these tissue components help pathologist to predict the type of cancer, it's staging and grading. Due to increased incidence rate, manual cancer diagnosis is tedious and can have a high chance of subjectivity, and is an inefficient process, making it unmanageable for large volumes of slides.

Computer-assisted diagnosis of cancer from histopathology images is drawing keen research interest in the medical community. Rapid development in high spatial resolution image acquisition and digitization techniques has lead to efficient digital data collection from the slides. However, the manual assessment of these slides even on a computer is a time consuming and subjective process. In this respect, artificial intelligence methods, especially the contemporary ones based on deep learning (e.g.

CNN models) could yield a promising solution that can assist pathologists to provide results that are timely and objective.

Computer aided automated analysis of breast histopathology images is challenging due to the existence of intra-tumor and inter tumor heterogeneity and inconsistency of stains in the tissue slides. Varieties in image scanners, different staining procedures and image acquisition protocols are some factors contributing to the inconsistencies in stains [2]. The stain variability make the extraction of relevant information from the slides particularly difficult for pathologist. Even though CNN based powerful algorithms are adopted for the classification of digital histopathology images, the fundamental changes in the input data feeding to CNN should be handled properly to deliver potentially improved results. Normalizing the stain using image processing and thus improving the data for deep learning model is often referred to as stain normalization. Number of stain normalizations techniques are reported in literature [2]-[5]. The stain normalization method map the stain colors of the analysed image by using the chromatic characteristics of specific reference image.

1.1 Contribution

In this work, we report a study for breast cancer histopathology image classification with two-fold objectives:

(1) we provide a performance review of some standard network models considering stain normalization, which is an integral part of the slide analysis pipeline. Here, we employ the method of Adaptive Color Deconvolution (ACD) for stain normalisation technique, and compare the performance of various CNN models with and without stain normalization.

(2) In choosing the CNN models we also consider the relatively less explored but, arguably, more compact CNN models such as MobileNet and EfficientNet, in addition to the VGG model, so as to study the trade off between performance and efficiency of the models. These pre-trained models are employed with some transfer learning for the current application. The proposed study involves extensive experiments on publicly available BreaKHis dataset which contains 7909, H&E stained breast histopathology images.

1.2 Organization of the paper

The rest of the paper is organized as follows. Section 2 reviews the literature, focusing on CNN models and stain normalization methods for histopathological image analysis. We present the methodology used in this work, for the stain normalization and classification of histopathological images in section3. Section 4 discusses experimental validation of the methods. Section 5 concludes with the results along with some discussions.

## 2. Related Work

Starting from some of the initial attempts reported in [6], to more recent works such as [7-11], conventional methods including image pre-processing, segmentation, feature extraction, and classification are adopted to develop automated diagnostic system for histopathology image analysis. The inception of Convolutional Neural Networks (CNN) enabled greater strides in efforts to improve the performance of breast cancer detection with more promising results [4]-[11]. For example, authors in [7], [12] present CNN models trained from scratch, designed for mitosis detection and Ductal Carcinoma In Situ classification. Availability of limited number of image samples make the usage of CNN difficult to achieve better performance for CAD systems. Ciresan et.al, in [12] propose the extraction of overlapping patches to improve the performance of CNN model with limited number of data. In [13], Spanhol et al, introduced a new dataset BreaKHis to the research community to explore the two class and multi-category classification of breast histopathology images.

Since we use BreaKHis database for our research and study, we report previous works from the literature that use the BreaKHis database for binary and multi class classification. Spanhol et.al, in [14], report the classification of images in the BreaKHis dataset with AlexNet. In this work, high-resolution random patches are extracted from the images for training. Yassir Benhammou et.al, present a review of existing deep net implementations on the BreaKHis dataset in their paper [15].

To improve the CAD system's performance for breast cancer detection, there have been many attempts to alter the layer distribution in existing networks. For example, in [16], authors adopt a new

Multiple Instance pooling layer in VGG Net to analyze the breast histopathology images. The pooling layers combine relevant features from image patches, without necessitating inter-patch overlap.

Jiamei Sun et.al, in [17] pursuant to a meticulous scrutiny into the efficiency comparison of the transfer learning technique, with training CNN from scratch, reaffirmed that transfer learning is more effective. The impact of the size of the context of training samples is evaluated in this work. Considering transfer learning, they make a comparison of ResNet 50, CaffeNet and GoogleNet.

Ziba Gandomkara.et.al in [18] introduce a Multi-class classification of breast cancer using deep residual Networks, which consists of two stages. The stage 1 is designed for classifying patches as benign and malignant using a deep residual network (ResNet). The stage 2 is used for four subclasses classification of benign and malignant tumor images.

In [19], Joke A. Badejo et.al, evaluate the performance of hand-designed and machine-designed descriptors. The experiments were validated using four datasets, including BreaKHis. Local texture descriptors designed with signal processing techniques (hand-designed) and features derived from the images by CNN (machine-designed) were analyzed.

In [20], Dalal Bardou et.al, compare two machine learning methods for the two class and multi class classification of breast histology images. The performance of feature extraction using CNN is compared with a set of handcrafted features derived by using bag of words and locality constrained linear coding. The performance is improved when they apply the ensemble model and SVM. They also reported multi-class classification of BreaKHis images using ensemble model with SVM.

Duc My Vo et.al, in [21] mainly focus on developing an ensemble of Deep Convolutional Neural Networks (DCNNs) using inception networks adopted to extract both global and local features of breast histopathology images. Stain normalization and data augmentation are applied as a pre-processing step. The Inception-ResNet-v2 model at various input scales are used to obtain multi-level features of breast images. They propose a gradient boosting trees algorithm to enhance accuracy of the DCNN classifiers. The majority voting strategy is applied to ensemble the boosting trees classifiers.

Vibha Gupta et.al, in their paper [22] present a sequential framework that uses multi-layered deep features that are derived from DenseNet. Dimensionality reduction has been performed using the XG Boost technique.

In [23], authors design a CAD system for the breast cancer classification using four convnets such as ResNet-50, Inception_V3, Inception_ResNet_V2, Xception as feature extractors and Logistic Regression, SVM, KNN as classifiers.

J. Xie et.al, in [24] present supervised and unsupervised DCNN for breast cancer classification. They adopt Inception v3 and Inception ResNet v2 architectures for binary and multi-classification using transfer learning techniques.

Y. Jiang et.al, in [25] argue that a small SE-ResNet module reduce the number of network parameters to develop a CNN efficiently. A new learning rate scheduler is used to fine-tune the network. They used the approach for both binary and multi class classification of BreaKHis images.

M. B. H. Thuy et.al, in their paper [26] propose a network combining deep learning, transfer learning, and generative adversarial network (GAN) to improve the classification performance. They use the VGG16 network and then both VGG16 & VGG19 to extract the features.

S. Saxena et.al, in [27] investigate the performance of pre-trained CNNs as a feature extractor for breast histopathology image analysis. They experimented on ten different neural networks reported in the literature. The extracted feature sets are fed to linear SVM.

M. Gour.et.al, in [28] propose a ResHist CNN model for breast cancer histopathological image classification. The images are pre-processed using stain normalization, data augmentation, and patch extraction.

Anupama et.al in [29] propose a capsule network which captures spatial and oriental information to classify breast cancer. Bayramoglu et.al, in [41] introduces a magnification independent model for binary classification of breast cancer using BreaKHis dataset. They make a comparison between magnification specific and independent models and conclude the better efficiency of magnification independent model. They put a hand to develop a generic deep learning model for breast cancer detection.

Mesut Togacar et.al in [30] develops the BreastNet model which consists of convolutional block attention module (CBAM), dense block, residual block and hypercolumn technique. A comprehensive review of the CNN designed for BreaKHis dataset is published in [31]-[32].

Most of the work mentioned above adopted transfer learning techniques for classification of breast histopathology images. Starting with AlexNet as a base model, researchers worked to improve the accuracy of CNN by investigating the effect of network depth, and have further looked into employing deeper network structures such as VGGNet, GoogLeNet, ResNet etc. However, these advances to improve accuracy by increasing the depth, haven't considered the network's efficiency in respect of size and speed.

Apart from classification, there are also many attempts using data pre-processing techniques to enhance the quality of images and reduce the appearance variation among images in a dataset. These are typically termed as stain normalization methods which can be used the initial stages of the development of an efficient CAD system [5].

D. Onder et.al, in [3] use various color normalization techniques as pre-processing steps on histopathology images to overcome the stain variability. But they haven't reported the efficiency of the stain normalization technique on their models. The commonly used stain normalization techniques are based on [2][4][5].

Marc Macenko et.al, in [4] propose an algorithm for stain normalization based on optical density values of the colors present in the image. From which the matrices of the stain vectors and the saturation of each of the stains are calculated. The effectiveness of the algorithm is verified on H & E stained histopathology images.

Adnan Mujahid Khan et.al, in [5] present a separate transformation based color deconvolution algorithm to distinct stain vectors in the images. Pixels having similar stains classified, and a staining vector is measured by seeing the arrangements of the pixels belonging to the stain. A non-linear mapping approach is presented to map the image color to the reference image in accordance with the stain vector. Structure preserving capability of separate transformation based stain normalization techniques are weak compared to unified transformation-based methods.

Abhishek Vahadane et.al, in [2] presents an approach based on structure-preserving stain normalization of tissues in the histopathology image. The color normalization based on unified transformation outperforms the separate transformation-based methods. To achieve better efficiency, [33] presents an ACD algorithm-based color normalisation method that handles the color artifacts better by preserving the structure of histopathology images.

## 3. Methodology

Elucidating our approach, in this section we describe the following; firstly, image pre processing with ACD based stain normalization technique to address the stain variabilities in the breast histopathology images. Secondly, implementation of three standard CNN models, namely VGG16, MobileNet and EfficientNet, and comparing their performance with and without stain normalization; finally, we also compare the efficiency and performance of these, arguably, low-footprint networks with the well-known DenseNet (wherein the latter has been used for the BreakHis dataset.) Fig.1 depicts the flow diagram of the binary classification of breast histopathology images.

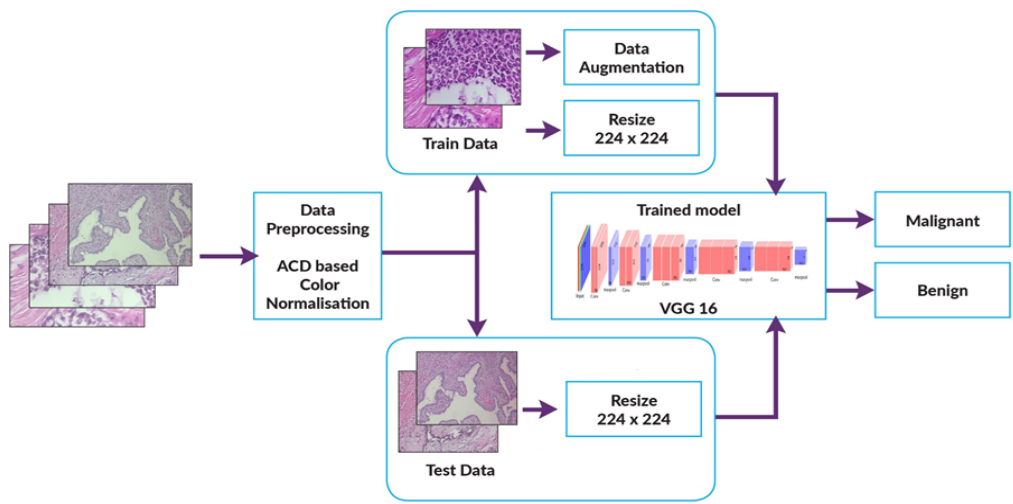

Fig.1: Overall flow diagram of the proposed model

### 3.1. Dataset Description

We evaluated our system with an openly available dataset called BreaKHis, made available by P & D Lab, Brazil [13]. The database consists of 7,909 images with a size of $700 \times 460 \times 3$. These images are obtained from 82 patients, comprising 2,480 benign, and 5,429 malignant samples. The dataset also consists of images of different magnifications such as $40\times$, $100\times$, $200\times$, and $400\times$. Benign class in the dataset includes Adenosis, Fibroadenoma, Phyllodes Tumor, and Tubular Adenoma. Malignant classes include Ductal Carcinoma, Lobular Carcinoma, Mucinous Carcinoma, and Papillary Carcinoma. However, here we consider the task of classification between 2 classes (Benign vs Malignant).

Primarily, via experiments we analyze the effect of stain normalization using images of 200× and 400× magnification factors with 2,013 and 1,820 samples, respectively.

### 3.2. Stain Normalization

We adopted Unified color transformation based stain normalization using ACD algorithm [33] and found this approach proved quite effective, although other competitive stain normalization methods can also be used. We extract the independent stain components from each pixel of the breast histopathology image and separate the stain details using unified transformation. H & E stained images obtained from BreaKHis database are sampled and converted into corresponding Optical Density (OD) values as per the following expression:-

$$od_i = -ln\left(\frac{I_{xi}}{I_m}\right) \qquad (1)$$

where, $od_i$ denotes Optical Density (OD) of RGB channels, $I_{xi}$ is the RGB values of the i[th] pixel in the image and $I_m$ is the intensity value of the background.

The Stain Color Appearance (SCA) matrix of image pixels is derived from OD converted RGB values. ACD matrix is derived by taking the inverse of the SCA matrix. A stain weight matrix W is defined, where W =diag($W_h$,$W_e$,1). H & E stain densities $s_i$ are obtained as shown in equation (2).

$$s_i = (W).(D).(od_i) \qquad (2)$$

where $s_i \in R^{3 \times 1}$

The obtained stain densities, $s_i$ are optimized with the help of an objective function shown in (3).

$$L = L_p + \lambda_b L_b + \lambda_e L_e \qquad (3)$$

Where $L_p$, $L_b$ and $L_e$ are obtained from equation (4),(5) and (6).

$$Lp = \frac{1}{N}\sum_{K=0}^{N} d_k^2 + \lambda_p \frac{1}{N}\sum_{k=1}^{N} \frac{2 h_k e_k}{h_k^2 + e_k^2} \qquad (4)$$

First term in (4) reduces the residual of the separation, second term represents the pixel values having H or E stain after the separation. N is the number of pixels used for the optimization and $p$ is the weight of the two stains factors in the image. λp is the weight of the first and second term in equation (4). Two more factors $L_b$ and $L_e$ are embedded in the $L$ function to handle the proportion of the two stains and overall intensity of the stains.

$$Lb = \left[(1-\eta)\frac{1}{N}\sum_{k=1}^{N}h_k - \eta\frac{1}{N}\sum_{k=1}^{N}e_i\right] \quad (5)$$

$$Le = \left[\gamma - \frac{1}{N}\sum_{k=1}^{N}h_k + e_k\right]^2 \quad (6)$$

where $\eta \in (0, 1)$ is considered as the balance parameter. Similarly, $\gamma$ is defined to balance the desired stain intensities. $\lambda b$ and $\lambda e$ are considered to be the weights.

The objective function in equation (3) is optimized using a gradient descent optimization algorithm for obtaining the adaptive variables for separation and normalization of H & E stains in the histopathology images. With these variables, H & E stain components of the images are separated. Finally, recombination of the weighted stain components with the SCA matrix of template images selected from BreaKHis dataset has been done to perform stain normalization.

### 3.3. Data Augmentation

To increase the amount of training data for the CNN models, we selected some data augmentation methods, whose parameters are mentioned in Table 1.

Table 1: Data Augmentation Parameters

| Parameter | Value |
| --- | --- |
| Shear range | 0.2 |
| Zoom range | 0.2 |
| Rotation range | 25º |
| Horizontal flip | True |
| Width shift range | 0.1 |
| Height shift range | 0.1 |
| Fill mode | Nearest Neighbour |

We used *shear transform* for randomly slanting the shape of the image. One axis is fixed and stretches the image at an angle of 0.2 for this work. Zoom is for zooming the image details and we use a zoom value of 0.2. Rotation considered here indicates upto 25 degree random rotation of images in the dataset. Horizontal flip is for flipping images horizontally. Width shift and height shift ranges measured as a fraction of total width or height of 0.1 by which the image is to be randomly shifted, either towards the left or right. Fill mode strategy of nearest is considered for filling in newly created pixels, which can appear after a rotation or a width/height shift

### 3.4. Deep learning methods

To classify stain normalized histopathology images, we adopted the lightweight VGG16, MobileNet, EfficientNet models, and compared the efficiency of these models with DenseNet which has already proven better performance for BreaKHis dataset. The image resolution is modified to 224 × 224 pixels to suit the requirement of input image sizes, conforming to the mentioned CNN models. VGG16 consists of thirteen convolutional layers and five maxpooling layers, with three fully connected layers stacked with convolution layers [34].

Andrew G. Howard et.al [35] introduced MobileNet, which is augmented by depthwise separable convolution unlike VGGNet, GoogLeNet etc, to reduce model size, complexity and make a lightweight CNN model suitable for mobile and embedded vision applications. Depthwise convolutions are applied to a single channel at a time, which reduces the number of parameters and makes MobileNet a low footprint model. The advantages of having fewer parameters in MobileNet, can also help to reduce overfitting in the model. Also, additional parameters such as width multiplier and resolution multiplier further enables it to be suitable for Mobile and embedded applications. In [35], the obtained accuracy on the ImageNet dataset is similar to that of VGGNet, but with lesser complexity. MobileNet is constituted of 28 layers, where a batch norm and ReLU nonlinearity follow all layers, except the final fully connected layer, then feeds into a softmax layer, for classification. Average pooling is added before the fully connected layer to reduce the spatial resolution to 1.

EfficientNet is a recently developed network, invented by Google that yields the property of model scaling that uniformly scales width, depth and resolution of the network using a highly effective compound coefficient [36]. A baseline network is first designed, by performing neural architecture search, for optimizing both the accuracy and efficiency that are measured based on the floating-point operations per second (FLOPS). EfficientNet uses the mobile inverted bottleneck convolution (MBConv) and depthwise convolution operation to make the model computationally cheaper. Through the reduction of parameter size and FLOPS, EfficientNet claims better efficiency compared to extant deep nets. The EfficientNetB0 is composed of sixteen inverted residual blocks (mobile inverted bottleneck MBConv) having different settings, including squeeze & excitation blocks along with swish activation.

Gao Huang et al in [37] introduced DenseNet, which is capable of handling vanishing gradient problems in a deeper CNN that arise when input passes through many layers, to the end of the network. Since, it requires fewer parameters than traditional CNN models, it enhances the efficiency of DenseNet. The DenseNet architecture has a dense connectivity pattern to ensure maximum information flow between layers in the network, making the training easier. In contrast to ResNet, instead of combining features through summation, DenseNet uses the concatenation of features to increase variation in the input of subsequent layers to improve efficiency, in spite of its depth.

In our work, parameters of all layers, except the fully connected layers, are frozen and transfer learning is performed with histopathology images. A fully connected layer is modified with 256 hidden neurons and Rectified Linear Unit (ReLU) activation functions. The regularisation technique, early stopping, dropout with probability of 0.4 and batch normalisation are added to handle the bias-variance trade-off in a better way. The Inference layer has been modified to support the binary classification of breast histopathology slides. The performance of the CNN models is evaluated by popular metrics, Test Accuracy, Specificity, Sensitivity, F1 Score and ROC analysis.

## 4. Experimental Results

In this section, we discuss the impact of stain normalization for binary classification of histopatholoical images using transfer learning methods. We demonstrate the following:

- the outcome of ACD color normalization technique using BreaKHis dataset
- the comparison of the performance of contemporary CNN classifiers before and after stain normalization.
- the results comparing the efficiency of the mentioned CNN classifiers, analysing the computational complexity of the models.

Fig.2 depicts the image samples from the BreaKHis database before and after stain normalization. In this work, we consider the 200× and 400× magnified benign and malignant samples. The normalization method used in this research work is a semi-automatic stain normalization procedure where the reference images are chosen manually by analysing various stain components in the images.

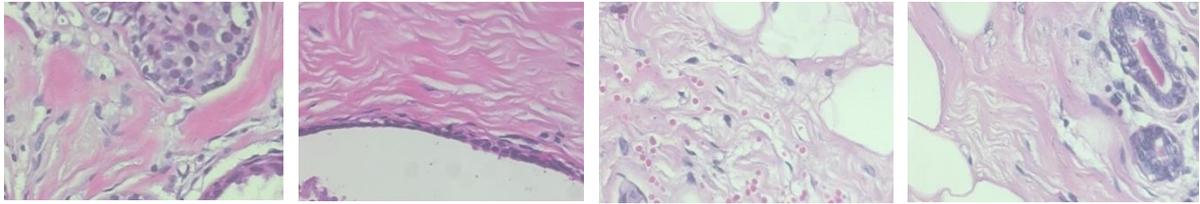

(a) 200× magnified benign samples

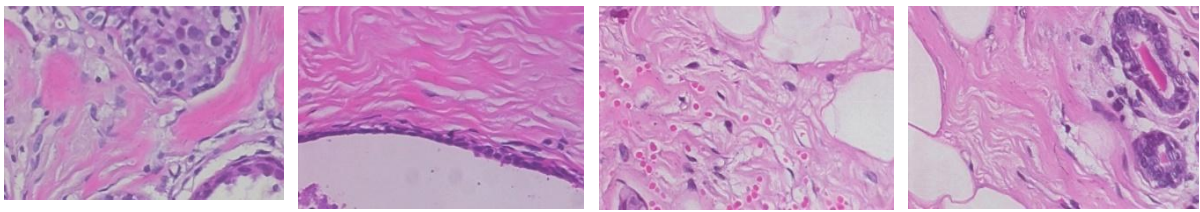

(b) image samples shown in Fig.2 (a) after stain normalization

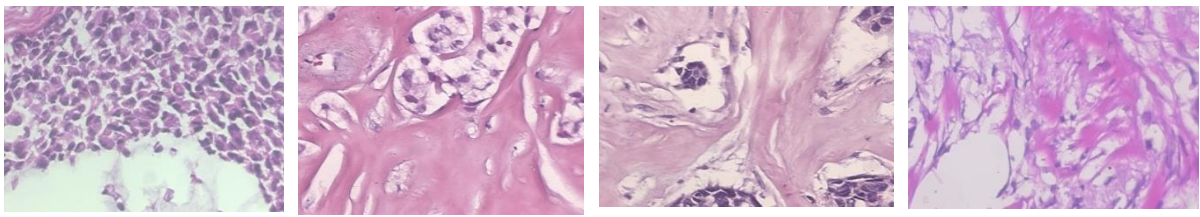

(c) 200× magnified malignant samples

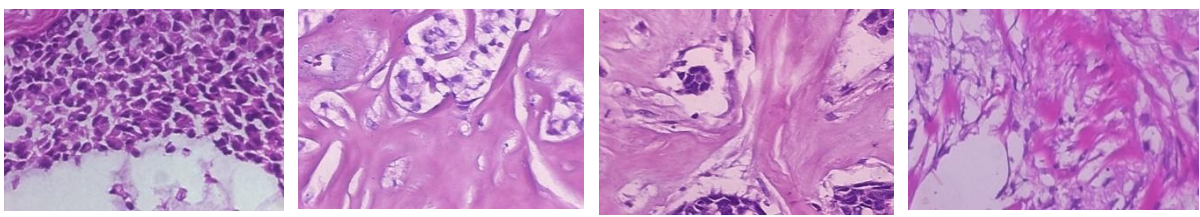

(d) malignant image samples shown in Fig.2 (c) after stain normalization

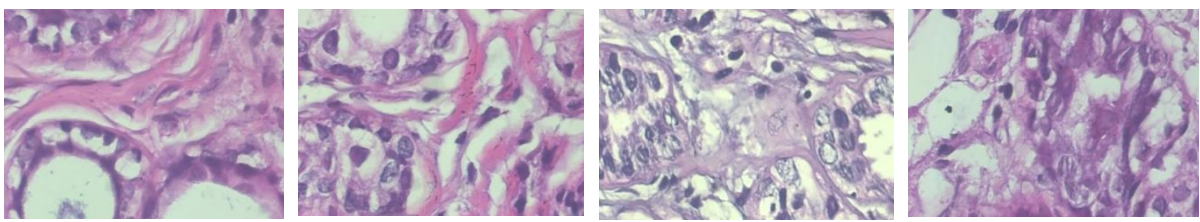

(e) 400× magnified benign samples

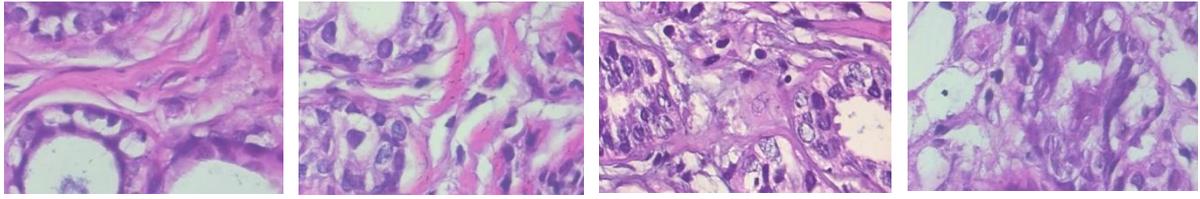

(f) benign image samples shown in Fig.2 (e) after stain normalization

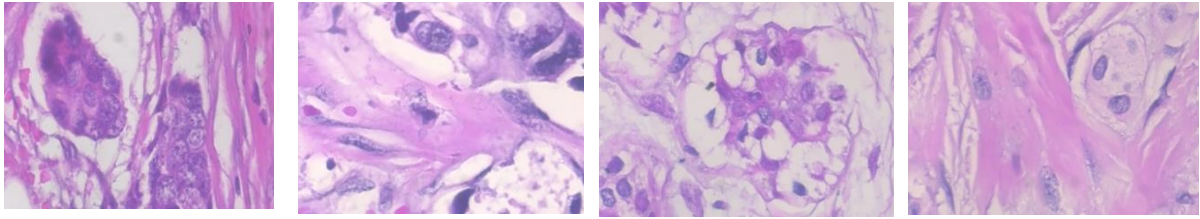

(g) 400× magnified malignant samples

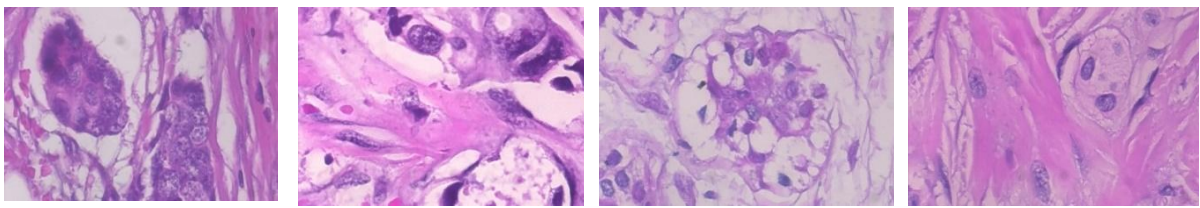

(h) malignant image samples shown in Fig.3 (g) after stain normalization

Fig 2: BreaKHis image samples with 200× and 400× magnification factors before and after stain normalization

Subjective analysis of the images reveal that after ACD based normalization, color variability is reduced and improvement in image contrast is also apparent

### 4.1. Experimental Setup

All images were resized to $224 \times 224$ pixels according to the input size of selected pretrained models. The batch size is set to 32 and all models trained for 100 epochs. A dropout layer with a probability of 0.4 is also used. We used Adam optimizer to optimize CNN weights for histopathology image classification. Learning rate decay method is adopted to reduce the learning rate, to ensure that the model achieves more stability at the later iterations. For fine tuning, we modified the last dense layer to output two classes corresponding to benign and malignant. The initial networks weights are set based on weights trained on ImageNet. A 70%-30% ratio is followed to partition the training and test data. Additionally, the training data is again split into training data and a validation data with a data split of 90 %-10 %. The validation dataset is used for monitoring the accuracy on the validation set and evaluation of model's performance. Training and testing process of the proposed architecture for this experiment is implemented in Python using Keras package with Tensorflow as the deep learning framework backend. The VGG16, MobileNet and EfficientNetB0 deep learning models are compiled with GPU support provided by Google Colaboratory, which permits a Tesla K-80 GPU instance, available for 12hrs. Google Colab is a collaborative version of the Jupyter / iPython notebook-based editing environment [38]. For the assessment of a model's performance, the experimental results derived from the CNN models are evaluated by metrics, such as accuracy, sensitivity, specificity and F1-Score.

### 4.2. Classification Results

This section presents the analysis of the effect of data augmentation and stain normalization for the classification of histopathology images.

*4.2.1 Performance comparison of of CNN models with and without stain normalization*

Table 2 and Table 3 summarize the effect of stain normalization for the performance of CNN models using data augmented BreaKHis images.

Table 2: Results comparison of stain normalization using 200× magnified images

| CNN Model | Magnification Factor | Preprocessing | Test Accuracy | Specificity | Sensitivity | F1 score |
|---|---|---|---|---|---|---|
| VGG16 | 200× | Without stain normalization | 0.9173 | 0.9367 | 0.8736 | 0.9401 |
|  |  | With stain normalization | 0.9428 | 0.9356 | 0.9625 | 0.9599 |
| MobileNet | 200× | Without stain normalization | 0.9595 | 0.9610 | 0.9121 | 0.9611 |
|  |  | With stain normalization | 0.9630 | 0.9816 | 0.9063 | 0.9737 |
| EfficientNetB0 | 200× | Without stain normalization | 0.9460 | 0.9465 | 0.9615 | 0.9641 |
|  |  | With stain normalization | 0.9479 | 0.9310 | 0.9938 | 0.9631 |

As reflected in Table 2, CNN models performed better with stain normalization for the classification of 200× magnified images. Stain normalized images produced a better score for evaluation metrics-test accuracy and F1 score. We did notice a slight drop in "specificity" value in the case of VGG16 and EfficientNet. But the "sensitivity" is significantly increased. Considering the three CNN models, we obtained the best test accuracy, specificity and F1 score with MobileNet.

We have tabulated the performance of CNN models using 400× magnified images in Table 3. We observed in this case too that the stain normalized images facilitate improved test scores. EfficientNetB0 exhibited better performance than VGG16 and MobileNet in terms of evaluation metrics- test accuracy, specificity and F1 Score by 0.54 %, 3.25 %, and 0.33 % respectively. However we noticed that sensitivity dropped by 4.2 %. However, the overall F1 score is better in case of EfficientNet.

Table 3: Results comparison of stain normalization using 400× magnified images

| CNN Model | Magnification Factor | Preprocessing | Test Accuracy | Specificity | Sensitivity | F1 score |
|---|---|---|---|---|---|---|
| VGG16 | 400× | Without stain normalization | 0.9433 | 0.9357 | 0.9682 | 0.9585 |
| | | With stain normalization | 0.9195 | 0.9945 | 0.7440 | 0.9418 |
| MobileNet | 400× | Without stain normalization | 0.9013 | 0.9078 | 0.8042 | 0.9028 |
| | | With stain normalization | 0.9307 | 0.9590 | 0.8333 | 0.9423 |
| EfficientNetB0 | 400× | Without stain normalization | 0.9470 | 0.9553 | 0.9312 | 0.9593 |
| | | With stain normalization | 0.9457 | 0.9754 | 0.8929 | 0.9636 |

Fig.3, Fig.4 and Fig.5 display a graphical comparison of the values tabulated in Table 5. We submit that as in the case of 200x images, it is consistent that stain normalized images produced better results for 400x images too, with EfficientNet proving to be more reliable.

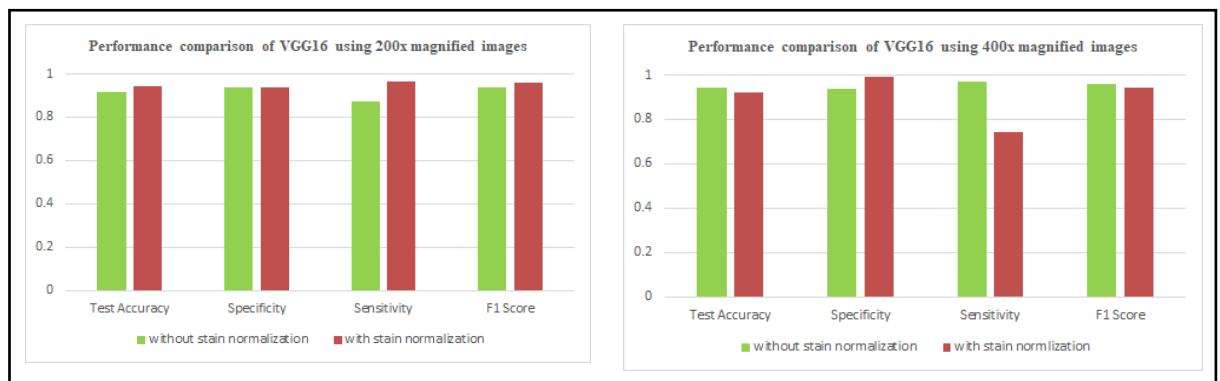

Fig.3: Performance comparison of VGG 16 model on images with and without stain normalization

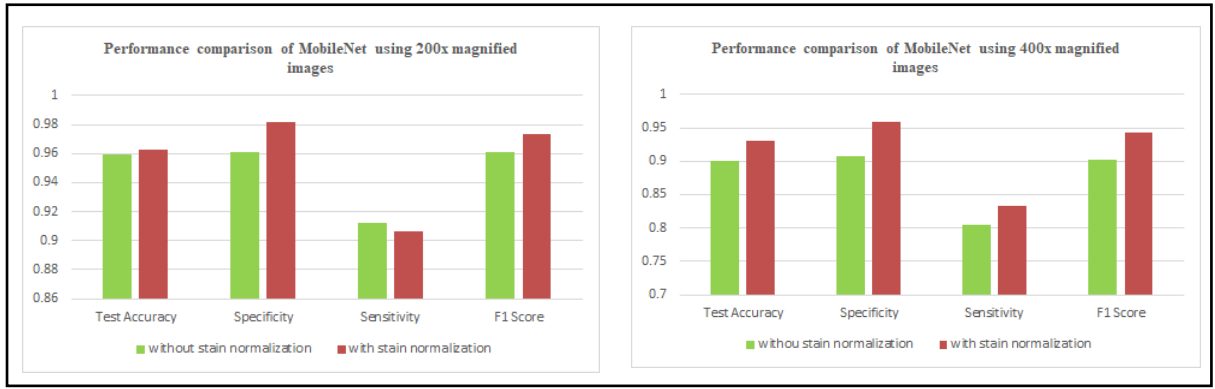

Fig.4: Performance comparison of MobileNet model on images with and without stain normalization

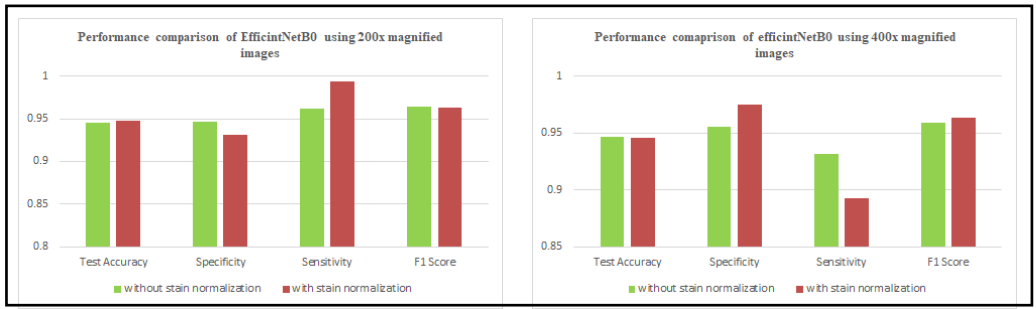

Fig.5: Performance comparison of EfficientNetB0 model on images with and without stain normalisation

*4.2.4 Results comparison of CNN model using stain normalized images in BreaKHis dataset with and without data augmentation*

The propitious effect of data augmentation on the training of histopathology images, was evident by a notably improved performance of the model. Fig.6, Fig.7 and Fig.8 depict the impact of data augmentation for training the stain normalized BreaKHis images. As in the above cases, we evaluated the performance of CNN models in terms of test accuracy, sensitivity, specificity and F1 score. Data augmentation consistently produced improved results for all the models, thus yielding a more reliable system for detecting malignancy, by better classifying the stain normalized images.

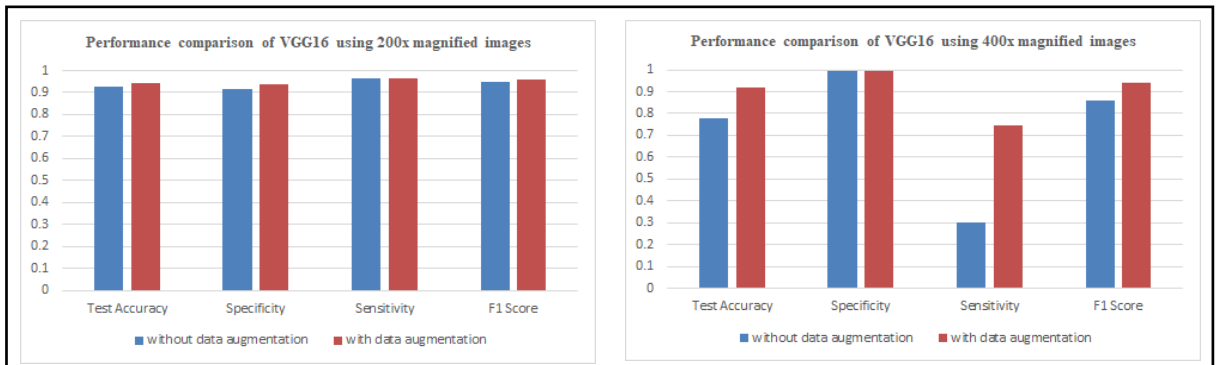

Fig.6: Performance comparison of VGG Net model on images with and without data augmentation

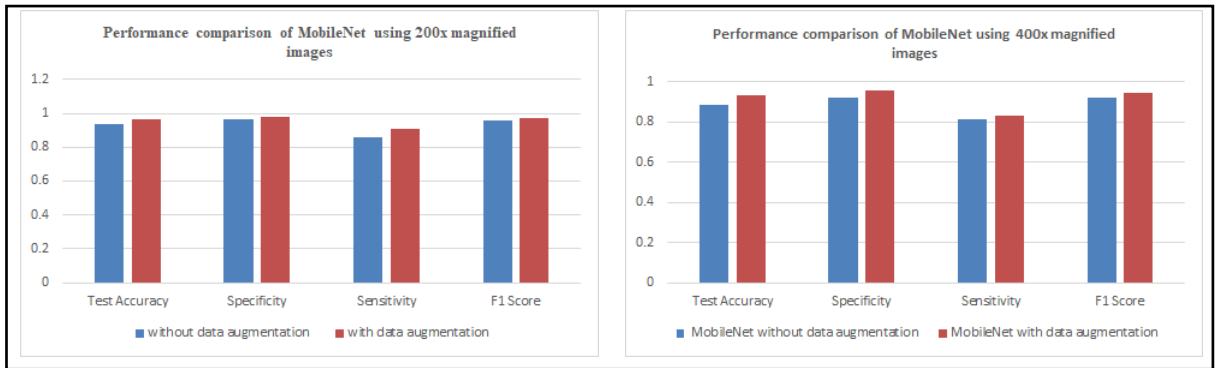

Fig.7: Performance comparison of MobileNet model on images with and without data augmentation

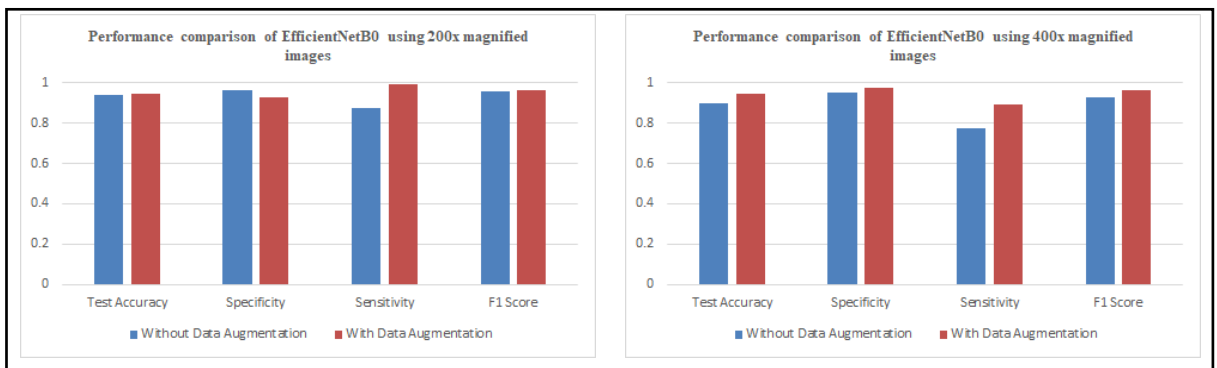

Fig.8: Performance comparison of EfficientNetB0 model on images with and without data augmentation

## 4.3. ROC Analysis

ROC analysis of classification model using VGG16, MobileNet and EfficientNetB0, before and after stain normalization on 200× magnified images are reported in Fig.9, Fig.10 and Fig.11. ROC curve was composed by accounting all possible combinations of True Positive Rate (TPR) and False Positive Rate (FPR), which show trade-off between sensitivity and specificity. From Fig.10 and Fig.11, it is clear that Area Under Curve (AUC) increases after we train the models with stain normalized images.

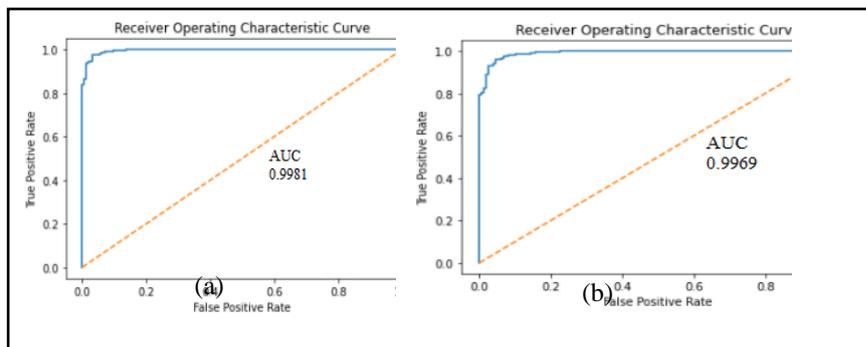

Fig 9: Plot of ROC curve using VGG 16 on (a) 200× magnified images without stain normalisation (b) 200× magnified images with stain normalisation

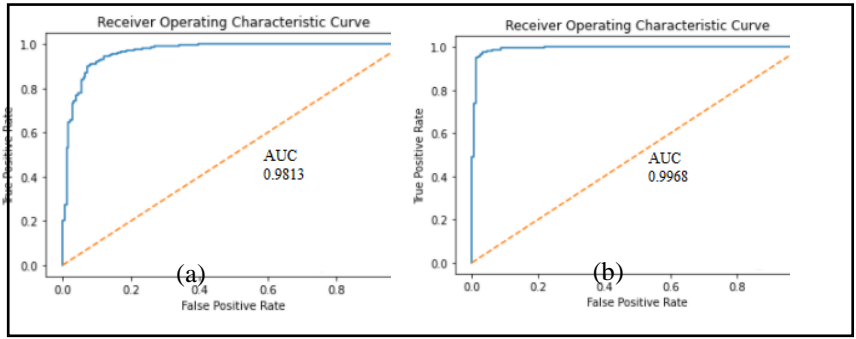

Fig 10: Plot of ROC curve using MobileNet on (a) 200× magnified images without stain normalisation (b) 200× magnified images with stain normalisation

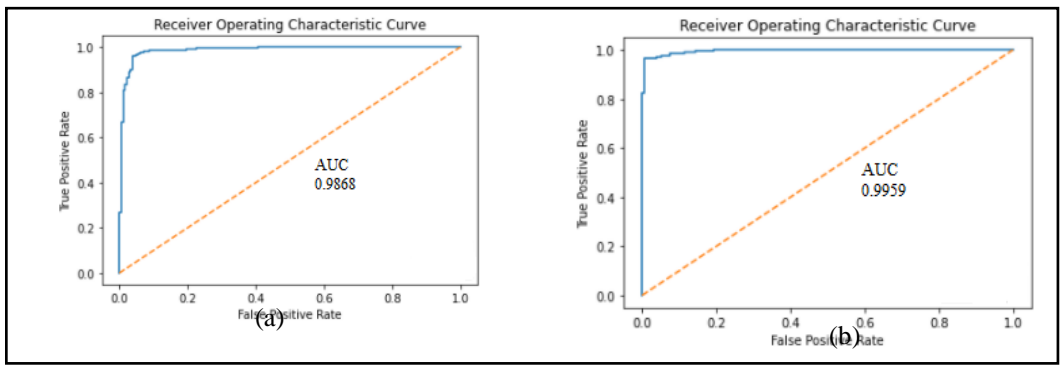

Fig 11: Plot of ROC curve using EfficientNetB0 on (a) 200× magnified images without stain normalisation (b) 200× magnified images with stain normalisation

The AUC of MobileNet and EfficientNetB0 on stain normalised images are found to be 0.9968 and 0.9959 respectively, illustrating the clear distinction between benign and malignant samples. However we notice that the VGG model performed well even without stain normalization. However, after stain normalization, all the lightweight models perform similarly in terms of AUC. This indicates the reliability of using stain normalized inputs.

Comparing the computational efficiency of the three CNN models, in terms of test time is tabulated in Table 4. Test time is defined as the time taken for predicting images in the test data set. 595 test images with 200× and 400× magnification factors were considered for performance evaluation. EfficientNetB0 proved to be computationally effective among the lot, consuming 123 seconds and 124 seconds for 200x and 400x images respectively. MobileNet and VGG16 models produced similar results with VGG16 having a slight edge over MobileNet.

Table 4: Efficiency comparison of VGG Net, MobileNet and EfficientNetB0

| Network | magnification factor | Test time (seconds) |
|---|---|---|
| VGG | 200× | 180 |
| VGG | 400× | 212 |
| MobileNet | 200× | 211 |
| MobileNet | 400× | 218 |
| EfficientNetB0 | 200× | 123 |
| EfficientNetB0 | 400× | 124 |

In addition to this, we compared the efficiency of EfficientNetB0 with DenseNet169 which has been established as an effective model in the literature for breast histopathology image analysis. We have tabulated the comparison results in Table.5. DenseNet is substantially deeper effectively reduces the number of parameters in the network and ensures maximum information flow between layers [37]. With fewer parameters, and deeper layers, DenseNet has been tested and proven for lesser computational complexity. However, in our experiments, EfficientNetB0 reported better computational test time and exhibited lesser complexity compared to DenseNet as shown in Table 5.

Table 5: Performance and efficiency comparison of DenseNet and EfficientNet

| Trained Model | Magnification factor | Test Accuracy | Specificity | Sensitivity | Precision | F1 Score | Test Time |
|---|---|---|---|---|---|---|---|
| DenseNet | 200× | 0.9482 | 0.9954 | 0.9500 | 0.9870 | 0.9709 | 204 |
| DenseNet | 400× | 0.9475 | 0.9590 | 0.9226 | 0.9112 | 0.9616 | 189 |
| EfficientNetB0 | 200× | 0.9479 | 0.9310 | 0.9500 | 0.8413 | 0.9631 | 123 |
| EfficientNetB0 | 400× | 0.9457 | 0.9754 | 0.8929 | 0.9434 | 0.9636 | 124 |

*4.4. Summary & Discussions*

This section provides in detail discussion about impact of stain normalization and classification performances through our approach. We selected three standard CNN models VGG16, MobileNet and EfficientNetB0 to analyse improvement in performance when we introduce ACD based color normalization and data augmentation methods for data pre processing. The processed images after stain normalization are as shown in Fig. 2. It is apparent that the stain variability is reduced and image contrast improves after stain normalization. We reduced the size of the processed images to 224× 224 to work with the three mentioned CNN models. Firstly, we evaluated the classification performance of the CNN models by using images in the BreaKHis database with and without data augmentation. For the experimental study, we used images with 200× and 400× magnification factors in the dataset. The analysis showed that the metrics such as classification accuracy, sensitivity, specificity, F1 score are improved when we added data augmentation during training the images. We submit that, increasing the number of images through data augmentation increases the performance of CNN models. Data augmentation can be applied both in online and offline mode. In our work, we used online data augmentation using ImageDataGenerator in the Keras deep learning library.

Secondly, we analysed the performance of CNN model using stain normalized and data augmented images. MobileNet and EfficientNet performed better after stain normalization. Almost all metrics improved after stain normalization. We noted that there is slight drop in specificity and sensitivity after stain normalization. We believe that more appropriate method of choosing reference images can help us to improve the sensitivity and specificity.

Finally, we analyse the efficiency of VGG16, MobileNet and EfficientNetB0 in terms of test time to verify the computational complexity of the CNN models. EfficientNetB0 outperformed VGG16 and MobileNet and DenseNet in terms of computational complexity

From the analysis we can say that every model has some advantages and some drawbacks. Among the considered CNN models, on an average we suggest that EfficientNetB0 is the fastest, has better performance accuracy and computationally efficiency. So this model can be used for devices with limited computational resources and low memory. DenseNet provides better accuracy, but needs more test time, and can be considered for the model implementation in a device with higher computational capability. The proposed models have some limitations in the case of sensitivity and specificity. This can be solved by considering more suitable reference images for the stain normalization. It should also be noted that an automated process of selecting reference images is yet to be achieved. Thus, manual intervention is still required to analyze histopathology images and automated data pre-processing techniques are called for.

**Experimental Comparisons**

list of various state-of-the-art research works, undertaken for the classification of breast histopathology images were summarized in section 2. Table 6 lists select models along with the F1-score attained by them. We may not be able to draw a completely fair comparison with the listed system since the training and test data split may be different in each case. Thus we shortlisted some CNN models that were reporting F1 score on BreaKHis dataset in the literature, and compared them with our approach. That way our results are compared based on the same dataset. Our primary goal was to experiment the effectiveness of stain normalization on classification with lightweight network. From our analysis, we noted that such networks perform similar to the other sophisticated models such as ResNet, GoogleNet, CaffeNet [14], and some ensemble models.

Table6: Comparison of the proposed model with state-of-the-art

| Paper | Methods | F1 Score (%) |
|---|---|---|
| Spanhol et al [14] | AlexNet | 88.7(200x), 85.9(400x) |
| Spanhol et al [39] | AlexNet | 88.7(200x), 86.7 (400x) |
| Dalal Bardou et.al [20] | ensemble model and SVM | 97.41(200x).97.1(400x) |
| J. Xie et.al [24] | Inception v3 and Inception ResNet v2 | 99.48(200x), 99.15(400x) |
| **Proposed System** | **EfficientNet** | **96.31 (200x), 96.36(400x)** |

5. **Conclusion**

Computerized classification of histopathology images plays a vital role in providing timely and better accurate results to patients. In developing a CAD system, stain variability in the histopathology images can degrade the performance of the system. To address this challenge, this paper analysed the impact of stain normalization in the automated classification of breast histopathology images. Adaptive Color Deconvolution (ACD) based stain normalization algorithm was employed to mitigate the stain variability and color artifact in the H&E stained histopathological images. The stain normalized images were fed to CNN model such as VGG16, MobileNet and EfficientNet for the binary classification of images. Metrics like Test Accuracy, Sensitivity, Specificity, F1 score and ROC were used for comparison. In addition to it, we evaluated the performance and efficiency of four CNN models- VGG16, MobileNet, EfficientNet and DenseNet. From our evaluation it was apparent that stain normalization was critical in attaining better accuracy in classifying histopathology images. With accuracy attained on par with the state-of-the-art models in the literature, we can submit that our system can reliably facilitate breast cancer detection by analysing biopsy images. As future work, devising mechanisms to select most appropriate template images for stain normalization can enhance the whole process significantly. To train and evaluate the system based on multiple datasets will enable better evaluation of the effectiveness of the classification process.

The breast cancer detection by analysing biopsy images demonstrated in this work will thereby aid in developing a reliable CAD system. As a future research direction, most appropriate template images can be chosen for stain normalization and investigated effectiveness of transfer learning using different datasets and recently published CNN models.

**Acknowledgement**

We would like to express our sincere gratitude to our beloved Chancellor Dr. Mata Amritanandamayi Devi, popularly known as Amma, for the immeasurable motivation and guidance for the accomplishment of this work.

.